\documentclass[aps,prl,reprint,superscriptaddress]{revtex4-1}

\usepackage{graphicx}
\usepackage{microtype}
\usepackage{xcolor}
\usepackage{nicefrac}
\usepackage{amsmath}
\usepackage{amssymb}
\usepackage{gensymb}
\usepackage{ulem}
\usepackage{stix}
\usepackage[T1]{fontenc}
\usepackage[colorlinks=true, linkcolor=blue, anchorcolor=black, citecolor=blue, filecolor=black, urlcolor=blue]{hyperref}
\usepackage[all]{hypcap}

\newcommand{\ikf}{Institut f\"ur Kernphysik, J. W. Goethe-Universit\"at Frankfurt, Max-von-Laue-Str. 1, 60438 Frankfurt am Main, Germany}
\newcommand{\ESRF}{ESRF, The European Synchrotron, 71 Avenue des Martyrs, CS 40220, 38043 Grenoble Cedex 9, France}
\newcommand{\kassel}{Institut f\"ur Physik und CINSaT, Universit\"at Kassel, Heinrich-Plett-Str.~40, 34132 Kassel, Germany}
\newcommand{\berlin}{Molecular Physics, Fritz-Haber-Institut der Max-Planck-Gesellschaft, Faradayweg 4-6, 14195 Berlin, Germany}
\newcommand{\HD}{Max-Planck-Institut f\"{u}r Kernphysik, Saupfercheckweg 1, 69117 Heidelberg, Germany}
\newcommand{\EXFEL}{European XFEL, Holzkoppel 4, 22869 Schenefeld, Germany}

\begin{document}
\noindent\makebox[\textwidth][l]{Authors' version of \href{https://link.aps.org/doi/10.1103/PhysRevLett.133.183002}{Physical Review Letters \textbf{133}, 183002 (2024)}}\\
\title {
\large
Role of the Coulomb Potential in Compton Scattering}
\author{N.~Melzer} \email{melzer@atom.uni-frankfurt.de} \address{\ikf}
\author{M.~Kircher} \address{\ikf}
\author{A.~Pier} \address{\ikf}
\author{L.~Kaiser} \address{\ikf}
\author{J.~Kruse} \address{\ikf}
\author{N.~Anders} \address{\ikf}
\author{J.~Stindl} \address{\ikf}
\author{L.~Sommerlad} \address{\ikf}
\author{D.~McGinnis} \address{\ikf}
\author{M.~Schmidt} \address{\ikf}
\author{L.~Nowak} \address{\ikf}
\author{A.~Kügler} \address{\ikf}
\author{I.~Dwojak} \address{\ikf}
\author{J. Drnec} \address{\ESRF}
\author{F.~Trinter} \address{\berlin}
\author{M.~S.~Sch\"offler} \address{\ikf}
\author{L.~Ph.~Schmidt}\address{\ikf}
\author{\firstname{N. M.}~\surname{Novikovskiy}} \address{\kassel}
\author{\firstname{Ph.~V}.~\surname{Demekhin}} \address{\kassel}
\author{T.~Jahnke} \address{\HD}\address{\EXFEL}
\author{R.~D\"orner} \email{doerner@atom.uni-frankfurt.de}
\address{\ikf}

\begin{abstract}
We report a fully differential study of ionization of the Ne L-shell by Compton scattering of $20$~keV photons. We find two physical mechanisms which modify the Compton-electron emission. Firstly, we observe scattering of the Compton electrons at their parent nucleus. Secondly, we find a distinct maximum in the electron momentum distribution close-to-zero momentum which we attribute to a focusing of the electrons by the Coulomb potential.    
\newline\\
DOI: \href{https://link.aps.org/doi/10.1103/PhysRevLett.133.183002}{10.1103/PhysRevLett.133.183002}

\end{abstract}
\maketitle
At high photon energies, once the cross section for photo-absorption decreases to a level of $10^{-24}~\mathrm{cm}^{2}$, Compton scattering takes over as the dominant light-driven ionization mechanism. On the text-book level, Compton scattering is treated as the binary scattering of a photon at a free electron at rest leading to electrons with a momentum given by the photon momentum transfer $\vec{Q} = \vec{k_\gamma} - \vec{k_{\gamma'}}$ where $\vec{k_\gamma}$ and $\vec{k_{\gamma'}}$ are the momenta of the incoming and the scattered photon.
With the early work of DuMond \cite{DuMond29}, this picture has been refined by adding the initial momentum of the electron in the bound state to the momentum balance. This leads to a momentum distribution of Compton electrons given by the Fourier transform of the initial-state orbital shifted by $\vec{Q}$ 
(termed the impulse approximation \cite{Cooperbook}). 
It involves a far-reaching approximation, namely the neglect of the influence of the ionic potential from which the electron escapes. 
The consequences of this approximation are mostly unexplored to this day. The reason for this is that it is the electron which is strongly influenced by the potential, while in most experiments on Compton scattering one detects the scattered photon for practical reasons \cite{Cooper1985}.

It is the purpose of the present Letter to elucidate the role of the ionic potential for Compton scattering by accessing the electron momentum and the ion momentum emerging in the process. 
We exemplify this for the case of Compton scattering of $E_\gamma=20$~keV ($k_\gamma=5.37$~a.u.) photons at the Ne~2s and~2p shells (ionization potentials of $I_p=48.5$~eV and $21.6$~eV, respectively). The experiment has been performed at beamline ID31 of the European Synchrotron Radiation Facility (ESRF) in Grenoble, France using the COLTRIMS technique \cite{Dorner2000,Ullrich2003}. Details on the experiment are listed in the appendix.
We compare our experimental results to calculations based on the $A^2$ approximation \cite{Pisk_2016,BERGSTROM19973}. Hereby, one vector-potential operator annihilates the incident photon, and another one creates the scattered photon. For a given photon momentum transfer $\vec{Q}$, the respective transition matrix element (to the lowest order of perturbation theory) reads:
\begin{equation}
\begin{split}
A_{\vec{p}i} (\vec{Q}) & =\langle \Psi^-_{\vec{p}} \vert e^{i\vec{Q}\cdot \vec{r}} \vert \Psi_i\rangle 
. \label{eq:theory}
\end{split}
\end{equation}
Here, $\Psi_i$ and  $\Psi^-_{\vec{p}}$ are the wave functions of the initial bound and final continuum states of the Compton electron, and energy conservation is fulfilled as $E_\gamma-E_{\gamma'}={I_P}+p^2/2$. 
The numerical calculations were performed in the Hartree-Fock approximation by the stationary single-center method \cite{10.1063/1.3526026,Demekhin2007,Novikovskiy_2022}, including partial electron continuum waves with angular momenta up to $\ell < 50$. The present theoretical approach goes beyond the impulse approximation in one decisive point: the employed final states are the continuum eigenfunctions of the singly charged Ne$^+$ ion, while the impulse approximation assumes free electrons described by plane waves. Therefore, the interaction of the Compton electron with the ion is implicitly included in our approach to all orders, while it is neglected in the impulse approximation.

\begin{figure}[!t]
\includegraphics[width=.90\columnwidth] {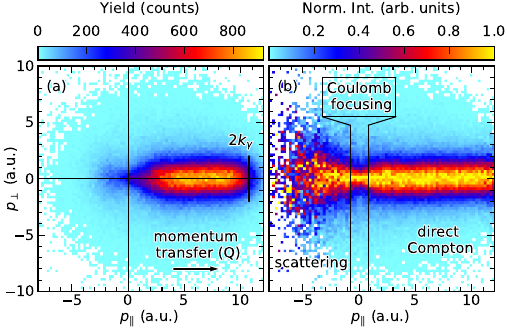}
\caption{Momentum distribution of Compton electrons recorded at $E_{\gamma}=20$~keV ($k_{\gamma}=5.37$~a.u.) for the Ne~2s and~2p shells. Horizontal axis: electron momentum parallel to the momentum transfer $\vec{Q}$ ($\|$). Vertical axis: electron momentum in one direction perpendicular to the momentum transfer $\vec{Q}$ ({$\perp$}). The data are integrated over the third electron momentum component and all photon momentum transfers. (a) The momentum distribution. The maximum momentum transfer (i.e., photon back-scattering) is approximately $2k_{\gamma}$, as indicated in the panel. (b) Same as panel (a) but each column of the histogram is normalized to its maximum. Three electron momentum regions dominated by different mechanisms are indicated (see text).}
\label{fig1}
\end{figure}

Figure~$1$ shows the electron momentum distribution measured for this process 
in a coordinate frame where the momentum transfer $\vec{Q}$ defines the horizontal axis. 
As expected, we find the electron being emitted mainly along the momentum transfer spanning from $0$ to $10.74$~a.u., the latter is the momentum transfer for photons of 20~keV back-scattering at a free electron at rest. According to the impulse approximation, the width of the distribution in the $\perp$ direction is given by the initial-state momentum distributions, which belong in the present case to the Ne~2s and~2p shells. A more detailed view of the electron momenta can be obtained when normalizing the two-dimensional distribution shown in Fig.~\ref{fig1}(a) for each column individually to the maximum value of that column
in  Fig.~\ref{fig1}(b).
The region of momenta where the measured distribution follows the behavior predicted by the impulse approximation is labeled ``direct Compton''. Contrary to this expectation, however, we find two additional contributions to the electron momentum distribution, one close-to-zero momentum, labeled ``Coulomb focusing'', and a second one in the backward direction, labeled ``scattering''. Electrons in this third region are emitted in the opposite direction of the momentum transfer. Electrons in the Coulomb focusing and the scattering regions are, as we will argue, a consequence of the interaction of the escaping Compton electron with the left-behind ionic potential.
As evidenced in the figure, the momentum distribution close to the origin is much narrower than that of the direct Compton electrons and, thus, significantly narrower than the initial bound-state momentum distributions of the Ne~2s and~2p shells. 
Before we analyze this observation in more detail, we discuss the backward-scattering region. The normalized distribution in Fig.~\ref{fig1}(b) shows that the momentum distribution of these backward-emitted electrons is significantly wider than the bound-state momentum distribution and the distribution of the forward-emitted direct Compton electrons. This is a direct indication of the mechanism responsible for the creation of these electrons. As they are emitted opposite to the momentum transfer, we argue that they are created by back-scattering of initially forward-kicked electrons at the ionic core. Then, the broad distribution in the $\perp$ direction is caused by the scattering at some angle other than $180\degree$. Thus, the observed large momenta perpendicular to the initial momentum transfer are a result of the deflection of this original momentum.

\begin{figure*}[tb]
\centering
\includegraphics[width=0.90\textwidth] {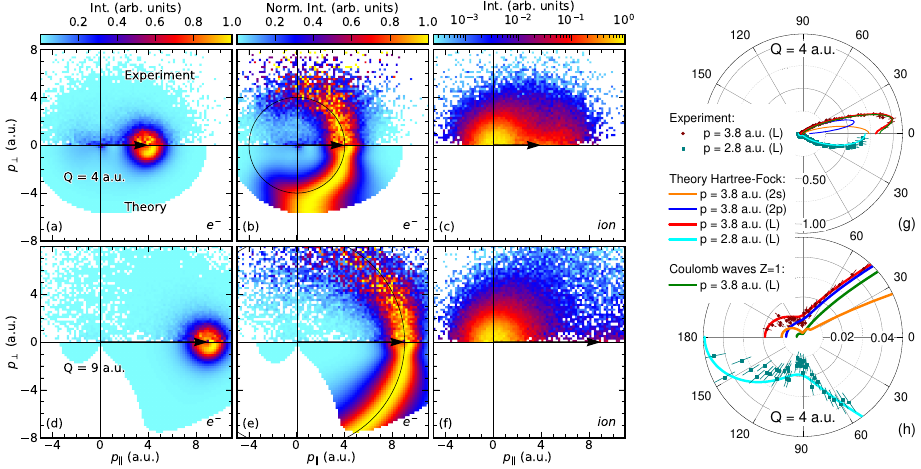}
\caption{Momentum distribution of Compton electrons (a,b,d,e) and ions (c,f) recorded at $E_\gamma=20$~keV ($k_\gamma=5.37$~a.u.) for the Ne~2s and~2p shells for fixed momentum transfer of $3.5~\mathrm{a.u.}<Q<4.5$~a.u. (a-c) and $8.5~\mathrm{a.u.}<Q<9.5$~a.u. (d-f), as indicated by the black arrows. Horizontal axis: electron (ion) momentum parallel to the momentum transfer ($p_\|$). Vertical axis: electron (ion) momentum perpendicular to the momentum transfer ($p_\perp$). The upper half of panels (a,b,d,e) show the experimental data, the lower half the theoretical results evaluating Eq.~(\ref{eq:theory}) in the $A^2$ approximation. (a,c,d,f) The color coding shows the differential cross section $d\sigma(p_\|,p_\perp)/dp_{\|} dp_\perp \cdot 1/p_\perp$. (b,e) The same as panels (a,d) but each row is normalized to its maximum value. The circles indicate the locus of events when the electron with the given momentum transfer is elastically scattered at the nucleus. (c,f) Counts are shown on a logarithmic scale. (g,h) Polar representation of the normalized emission probability distributions, obtained for photon momentum transfers $3.5~\mathrm{a.u.}<Q<4.5$~a.u. and electron momenta $p=3.8$~a.u. and 2.8~a.u. The experiment shows the results for the L-shell, the computed distributions are given separately for 2s, 2p electrons, and the combined L-shell. Results from Hartree-Fock calculations for Ne$^+$, as well as from modelling the continuum with Coulomb waves are depicted (see legend). All error bars indicate the standard statistical error. (g) Full distributions. (h) Distributions of panel (g), but enlarged by a factor 20 around the origin.}
\label{fig2}
\end{figure*}

To elucidate the three regimes of direct Compton, Coulomb focusing, and scattering in more detail, we show in Fig.~\ref{fig2} the electron momentum distribution for two different momentum transfers of $Q=4$~a.u. (top row) and $9$~a.u. (bottom row). If $Q$ was transferred to a free electron initially at rest, this would correspond to final electron energies of $218$~eV and  $1102$~eV, significantly larger than the binding energies of the Ne~2s and~2p shells ($I_p=48.5$~eV and $21.6$~eV). Both momentum transfers might thus be expected to be in the validity-regime of the impulse approximation. As expected, we find for both momentum transfers a distinct maximum in the electron momentum distribution centered at the momentum transfer vector [see Figs.~\ref{fig2}(a) and~\ref{fig2}(d)]. Closer inspection, however, shows in addition a narrow feature at zero electron momentum. This latter cusp-like distribution of slow electrons was labeled ``Coulomb focusing'' in Fig.~\ref{fig1}. This feature is also present in the theoretically calculated distribution. A third feature visible in Figs.~\ref{fig2}(a) and~\ref{fig2}(d) is a very broad low-count-rate distribution of events over an oval-shaped region (mainly in light blue). These events are neither electrons which are Coloumb-focused at the origin nor are they a pedestal of the $Q$-boosted initial-state momentum distribution. We argue that the underlying physics leading to electrons in this region of momentum space is elastic scattering of electrons at the nucleus. In Fig.~\ref{fig1}, such scattered electrons became visible in the backward direction.
{
The same process of scattering, however, does not only lead to back-reflection of the Compton-kicked electrons at the atomic potential but also to a deflection to all angles. This does not produce sharp structures in momentum space and thus is not easily seen. To render this scattering visible in Figs.~\ref{fig2}(b) and~\ref{fig2}(e) we normalize each row in Figs.~\ref{fig2}(a) and~\ref{fig2}(d) to the maximum in this row. This very intuitively shows the scattering to forward angles. Electrons with an initial momentum $\vec{Q}$ are elastically deflected at the nucleus, which distributes them along the sphere of radius $Q$ in momentum space, as highlighted by the circles. This elastic scattering is reproduced by our calculations using the $A^2$ approximation.
The intriguing angular dependent probability of this inneratomic scattering process is best seen in a polar representation [Figs.~\ref{fig2}(g) and \ref{fig2}(h)]. The scattering is highly sensitive to the exact shape of the potential, as a comparison of the backward-scattering peaks in Fig.~\ref{fig2}(h) for our full calculation using the Ne$^+$ Hartree-Fock potential (red curve) with the one from a Coulomb potential for charge $Z=1$ (green curve) shows. The dominating forward peak in Fig.~\ref{fig2}(g), caused by direct (i.e., unscattered) electrons, shows the contribution for the two electrons from the 2s (orange curve) and the six electrons from the 2p (blue curve) shells, with the structure of the 2p wave function in the total distribution (red curve) causing the dip at zero degree \cite{Ehrhardt_1974,Ren_2010}. This dip is only present for the binding-energy-corrected Bethe ridge at $p=(Q^2-2I_p)^{\nicefrac{1}{2}} \approx 3.8$~a.u. [see also Fig.~\ref{fig2}(a)]. The dip at $0\degree$ is also absent for smaller momenta, e.g.,  $p=2.8$~a.u. [cyan curve in Fig.~\ref{fig2}(g)], where the electrons are predominantly emitted from the 2s shell and, in addition, exhibit a stronger backward scattering peak 
[Fig.~\ref{fig2}(h)].

In an elastic-scattering event, the momentum is transferred to the nucleus. This is very different from the process of direct Compton scattering, where the nucleus is just a spectator \cite{Spielberger95} to the photon-electron momentum exchange. For the direct Compton scattering, the spectator-nucleus keeps the momentum it had in the bound state \cite{Kircher20natphys}. 
We present the measured Ne$^+$ ion momenta in Figs.~\ref{fig2}(c,f) using a logarithmic color scale. It shows the expected peak from the direct Compton mechanism close to the origin. In addition, it shows a significant amount of ions which are forward-emitted. A narrow contribution at the forward-momentum $Q$ is generated by the ions from the Coulomb-focusing mechanism. There are also ions more forward-shifted than $Q$. These are events in which the Compton electron is back-scattered at the nucleus, which leads to an ion forward-momentum of up to $2 Q$. There is also a significant forward tail of the momentum distribution at the origin. These are the ions related to the forward-scattering events emphasized in the two middle panels.

\begin{figure}[!t]
\includegraphics[width=.95\columnwidth] {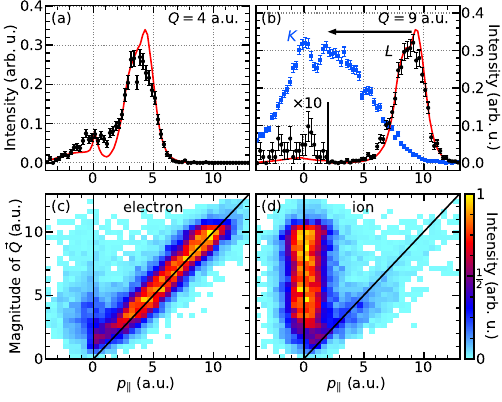}
\caption{Electron and ion momentum balance parallel to the momentum transfer for Compton scattering recorded at $E_\gamma=20$~keV ($k_\gamma=5.37$~a.u.) for the Ne $1s$ [panel (b) only], $2s$, and $2p$ shells. Horizontal axis: momentum parallel to the momentum transfer ($p_\|$). Vertical axis: intensity in panels (a,b), $Q$ in panels (c,d). Electron data are shown in panels (a,b,c), ion data in panel (d). Panel (a) shows momentum transfers where $3.5~\mathrm{a.u.}<Q<4.5$~a.u., panel (b) shows $8.5~\mathrm{a.u.}<Q<9.5$~a.u. (a,b) Black points show $2s$/$2p$-electron data; red lines corresponding theory calculations. All error bars show the standard statistical error. (b) Blue points show Ne $1s$-electron data. For small $p_{\|}$ the $2s$/$2p$-data is scaled by a factor $10$. See main text for an explanation of the arrow. (c,d) Vertical and diagonal lines indicate different mechanisms. The data is restricted to a momentum component perpendicular to $\vec{Q}$ of $\vert p_\perp\vert<0.4$~a.u.}
\label{fig3}
\end{figure}

We now come back in more detail to the feature labeled Coulomb focusing in Fig.~\ref{fig1}. In Fig.~\ref{fig2} it also manifests as a peak in the electron momentum distribution at the origin. To study its dependence on the momentum transfer, we select events with a momentum $\vert p_\perp\vert<0.4$~a.u. and inspect their momentum distribution along the momentum transfer in Fig.~\ref{fig3}. Panels (a) and (b) correspond to the two selected momentum transfers of $Q=4$~a.u. and $9$~a.u., respectively, just as in Fig.~\ref{fig2}. One recognizes the forward peak located at momentum transfers corresponding to the direct Compton process. As expected, this contribution is nicely reproduced by the calculations using the $A^2$ approximation. In addition to these direct Compton electrons, the Coulomb focusing produces a second, well-separated peak in the momentum distribution centered at zero, which was also seen in the two-dimensional momentum distributions of Fig.~\ref{fig1}. Also this peak is nicely reproduced by our calculations. With increasing momentum transfer, the intensity of the peak at zero reduces as compared to the contribution of the direct Compton peak. This can be directly seen in Fig.~\ref{fig3}(c) where the zero-peak fades out along the vertical axis. It is remarkable, however, that even at a momentum transfer as high as $9$~a.u., there are still electrons found which have close-to-zero momentum. The consequence of this surprising behavior can be seen in the ion momentum distribution as function of momentum transfer shown in Fig.~\ref{fig3}(d). Also here, one can recognize the ions with close-to-zero momentum originating from the direct Compton process. The Coulomb-focusing, however, produces a second group of ions which acquire the full momentum transfer. These are events where effectively the entire photon momentum is transferred to the nucleus, and the electron escapes with little momentum. Note that this finding is also reproduced by our calculations. These calculations do not involve, however, any nucleus or any scattering of photons at the nucleus. The matrix element (Eq.~\ref{eq:theory}) is simply the overlap integral of the momentum-boosted ground-state wave function with the continuum function from the respective ionic potential. The origin of the cusp at zero momentum can be revealed by  accounting for the binding energy in the impulse approximation (Fig.~\ref{fig4}). The $Q$-boosted initial-state momentum distribution has a small contribution close to the origin. Other than assumed by the impulse approximation, only those electrons with a kinetic energy exceeding the binding energy can escape. In order to model this classically, we subtract the momentum corresponding to the binding energy from the momentum of each electron while keeping its emission direction. The circle of radius $p=\sqrt{2I_p}$ in Fig.~\ref{fig4} indicates the escape boundary outside which electrons have a kinetic energy larger than the binding energy. The electrons' energy losses (when overcoming the binding energy) condense all electron density which is located just slightly outside of the escape boundary to pile up close to the origin [Fig.~\ref{fig4}(b)], forming the observed cusp. The quantum-mechanical analogues of this mechanism are automatically included in our calculations performed in the $A^2$ approximation, since the final continuum states are the eigenstates of the ionic potential rather than plane waves. To validate this physical explanation we compare in Fig.~\ref{fig3}(b) electrons emitted from the L-shell (binding energies 48.5 and 21.6~eV) to electrons emitted from the K-shell (binding energy 870~eV). As expected from our qualitative explanation, the amount of Coulomb-focused electrons at the origin increases drastically. In addition we see a shift of the position of the main peak (indicated by the black arrow) from the quasi free value $p=Q$ to the binding energy corrected value $p=(Q^2-2I_p)^{\nicefrac{1}{2}}$.

\begin{figure}[!t]
\includegraphics[width=.95\columnwidth] {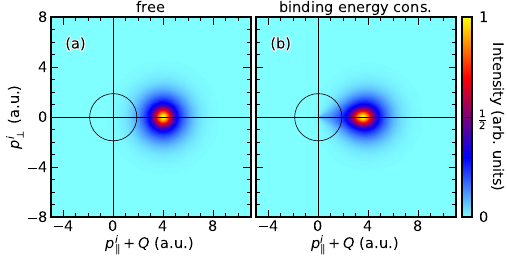}
\caption{Classical modeling of the Coulomb focusing. (a) Momentum distribution of Ne~2s and~2p shells shifted by a momentum transfer $Q=4$~a.u. The circle has a radius of $\sqrt{2I_p}$, thus only electrons outside this circle can overcome the binding energy. (b) For each electron in panel (a), the kinetic energy is reduced by the binding energy, while keeping the initial direction of the momentum vector. Electrons within the circle in panel (a) have, thus, a negative energy and are not shown in panel (b). Electrons on the surface of the sphere in panel (a) are focused to the origin in panel (b).}
\label{fig4}
\end{figure}

In conclusion, we have shown experimentally and theoretically that electron emission by Compton scattering is influenced by the ionic potential, leading to previously unexplored features in momentum space. This finding holds even at high photon energies and  momentum transfers, which correspond to energies that are high compared to the binding energy of the emitted electrons. The observed cusp-like contribution of electrons with close-to-zero momentum bares reminiscence to the target and projectile cusp, known in ion-atom collisions \cite{PhysRevA.58.2611}. Similar low-energy electrons are observed in tunnel-ionization processes in strong laser fields \cite{Blaga09}. Accordingly, we have adopted the term ``Coulomb focusing'' from strong-field ionization to describe them. 
We reported on a further contribution visible in our Compton spectra, which can be attributed to a scattering of the Compton electron at its parent ion. This process has analogies in other fields of atomic physics, as well. In (e,2e) collisions, it is this process which is responsible for the formation of the recoil peak in the angular distributions \cite{RevModPhys.66.985}. An analogous process has also been reported in ion atom collisions \cite{Bechthold.1997}. Along this line, studies on molecular inner-shell photoionization should be mentioned, where the photoelectron wave released from a specific site in a molecule is scattered at the neighboring atoms, illuminating the molecule from within \cite{Landers01}. All these related strong-field and photoionization processes point to the potential of our observation for molecular imaging. Our findings suggest that Compton scattering at inner-shell electrons in molecules can also be expected to yield distinct diffraction patterns, potentially complementing the established tools of laser-driven electron diffraction and photoelectron diffraction imaging.

\begin{acknowledgments}
This work is supported by the Deutsche Forschungsgemeinschaft (DFG). We acknowledge the European Synchrotron Radiation Facility (ESRF) for provision of synchrotron radiation facilities under proposal number CH-6729 (\href{https://doi.org/10.15151/ESRF-ES-1299363304 }{DOI: 10.15151/ESRF-ES-1299363304}) and we would like to thank V.~Honkimäki, H.~Isern, and F.~Russello for assistance and support in using beamline ID31. F.T. acknowledges funding by the Deutsche Forschungsgemeinschaft (DFG, German Research Foundation) - Project 509471550, Emmy Noether Programme.
\end{acknowledgments}

\textit{Appendix on the experimental details:}
Photons with a bandwidth of about $2\%$ were selected from the undulator beam using a pinhole monochromator \cite{Vaughan:kv5084}. The resulting photon beam with an intensity of around $2.6\cdot10^{14}$~photons/s was crossed with a supersonic Ne gas-jet target. All ions and electrons with an energy below $2$~keV where guided by parallel $50$~V/cm electric and $31$~Gauss magnetic fields onto two time- and position-sensitive detectors with hexagonal delay-line readout \cite{1046770}. An electron detector of $151$~mm active diameter was placed $24$~mm off center to accommodate the high-energy Compton electrons emitted to the forward direction. The ion arm of the spectrometer had a total length of $840.4$~mm and included an electrostatic lens \cite{DORNER1997225} to obtain the required ion momentum resolution of around $0.7$~a.u. We achieved a coincidence count rate of around $85$~Hz Compton events where the electron and the ion where detected. At $E_{\gamma}=20$~keV, the photoabsorption cross section of the Ne 1s shell is about a factor of $60$ larger than the cross section of the sought-after ionization by Compton scattering. The ions created by photoabsorption at the Ne 1s shell have a recoil momentum of $37.85$~a.u. from the emitted photoelectron and could be clearly separated from ions created by Compton scattering with typical momenta below $10$~a.u. (see, e.g., Fig.~1 in Ref.~\cite{Spielberger95} and Refs.~\cite{Samson94,Dunford04}).

\bibliographystyle{apsrev4-1}

\end{document}